# Femtosecond all-optical polarization manipulation of harmonic generation in an anisotropic strongly coupled plasmon-epsilon-near-zero metasurface

Yingying Zhang, Jinhai Si, Xun Hou, Kuidong Wang*

*Key Laboratory for Physical Electronics and Devices of the Ministry of Education and Shaanxi Key Lab of Information Photonic Technique, School of Electronic Science and Engineering, Xi'an Jiaotong University, Xi'an, 710049, China*

Email: wangkuidong@xjtu.edu.cn

## Abstract

Ultrafast manipulation of light polarization plays a key role in modern high-speed photonics and optoelectronic operations. However, most current polarization-active modulation schemes are restricted by a limited operating wavelength range (visible to infrared), narrow-band resonant fundamental wavelength working condition, and relatively long modulation time (> 1 picosecond). Here, by taking advantage of the transient anisotropic nonlinearity of optical strong coupling between the epsilon-near-zero mode of indium tin oxide (ITO) film and polarization-selective plasmonic resonance, we achieve femtosecond polarization manipulation of broadband light that spans ultraviolet (UV) to visible region by controlling harmonic generation of the signal light under resonant pumping. We demonstrate polarization elliptic rotation of 23.3° at 407 nm for third-harmonic generation (THG) and 55.0° at 610 nm for second-harmonic generation (SHG) simultaneously within ~600 fs. In addition, the strongly coupled system can also serve as a universal femtosecond polarization modulation source, imprinting a 32.9° polarization rotation in the UV region by converting the modulation information of the fundamental signal light to that of THG light via a uniform ITO film in a cascaded configuration. Our study establishes a new paradigm for UV-based, ultrahigh-speed broadband polarization photonics with direct implications for high-speed optical information processing and programmable nonlinear photonic devices.

# Introduction

Ultrafast active control of light polarization plays a crucial role in modern optical modulation and high-speed photonics. It serves as one of the bases for applications in high-speed optical imaging, integrated photonic computing and ultrafast lasers[1]. The realization of active polarization modulation relies on the anisotropic nonlinear optical response of the passing medium, whereas ultrafast modulation usually requires an all-optical control scheme along with the transient intrinsic nonlinear optical response of the material. To date, high-contrast, ultrafast control of light polarization has been implemented across diverse platforms, including laser-induced transient metamaterial[2], plasmonic metamaterials[3-7], dielectric meta-mirrors[8,9] and coupled epsilon-near-zero (ENZ) materials[10,11]. However, the operation wavelength of these modulation strategies is limited to visible and infrared ranges. Transient control of light polarization at shorter wavelengths, e.g., the ultraviolet (UV) region, is strongly restricted by a scarcity of natural nonlinear optical materials that exhibit both large nonlinearity and fast optical response at these wavelengths[12]. Furthermore, the current high-speed polarization modulations are usually achieved at the resonant fundamental wavelength of the materials, which typically rely on narrow-band optical resonances. This precludes the realization of broadband and multiple-wavelength polarization control in a single platform. Moreover, the response time of most reported polarization modulations is longer than picoseconds, which hinders the realization and application of ultrahigh-speed light polarization modulators and polarization-related photonic circuits operating beyond terahertz bandwidth (femtosecond response).

Optical strong coupling (OSC) has appeared to be a promising platform for femtosecond optical manipulation by combining large third-order optical nonlinearity with femtosecond optical response at the coupled resonances[13-15]. This coupled system occurs when the coherent energy exchange between two different optical modes exceeds each of their dissipation rates, resulting in an anti-crossing behavior in the dispersion relation and leading to the formation of new hybrid states[16,17]. The hybridization of the OSC with large coupling strength requires a small, confined volume of optical fields, resulting in strong local electromagnetic fields and further enhancement of nonlinear optical effects. Besides, the energy transfer between the transitions and cavity modes under strong coupling can experience Rabi oscillations with extremely fast response, e.g., on the order of tens of femtoseconds[18], which ensures the femtosecond response of the coupled system. Making use of these properties, OSC systems also exhibit efficient nonlinear wave-mixing effects, such as harmonic generation[19-21], sum-frequency generation[22], and four-wave mixing[23]. These wave-mixing effects extend the operating wavelength range of OSC to shorter wavelengths, while facilitating the realization of broadband and multiple wavelengths. Therefore, the unique nonlinear optical properties of OSC shown above make broadband, femtosecond optical modulation at shorter wavelengths promising.

Here, with the aid of strong harmonic generation of anisotropic strong coupling, we extend the operating wavelength of the light polarization modulation to UV region and investigate a broadband, multi-wavelength optical polarization control within hundreds of femtoseconds. The strong anisotropic optical nonlinearity for polarization modulation is realized by coupling a polarization-sensitive localized plasmonic mode to an ENZ mode. We achieve a polarization ellipse rotation of 23.3° and 55.0° within 600 fs for second-harmonic generation (SHG) and third-harmonic generation (THG), respectively. Such a transient modification of the polarization state of harmonic generation arises from hot electrons with different thermalized temperatures at various polarizations. In addition, we also design a cascaded femtosecond polarization modulation of UV light by converting the transient polarization modification of the near-infrared (NIR) fundamental light to that of harmonic generation. The strategy and design for achieving femtosecond polarization modulation in the UV region in this work offer a new platform for UV-based, ultrahigh-speed broadband polarization modulators and polarization-related photonic devices.

# Results and Discussion

## Design principle

The anisotropic strong coupling for light polarization manipulation was implemented by integrating a polarization-selective plasmonic mode of gold nanoantennas with the ENZ mode of an indium tin oxide (ITO) film. In this case, optical strong coupling only exists at a specific polarization direction, i.e., horizontal polarization ($x$-pol.) (**FIG. 1**). This ensures a polarization-dependent nonlinear optical response for the coupled system. To meet the requirements of polarization modulation of harmonic generation, the coupled system should generate measurable harmonic fields (both SHG and THG) along orthogonal polarization directions with the excitation of fundamental-wave (FW) signal light. This facilitates the characterization of Stokes parameters for the complete polarization state of harmonic fields and becomes critical when considering the change of polarization azimuth angle[24].

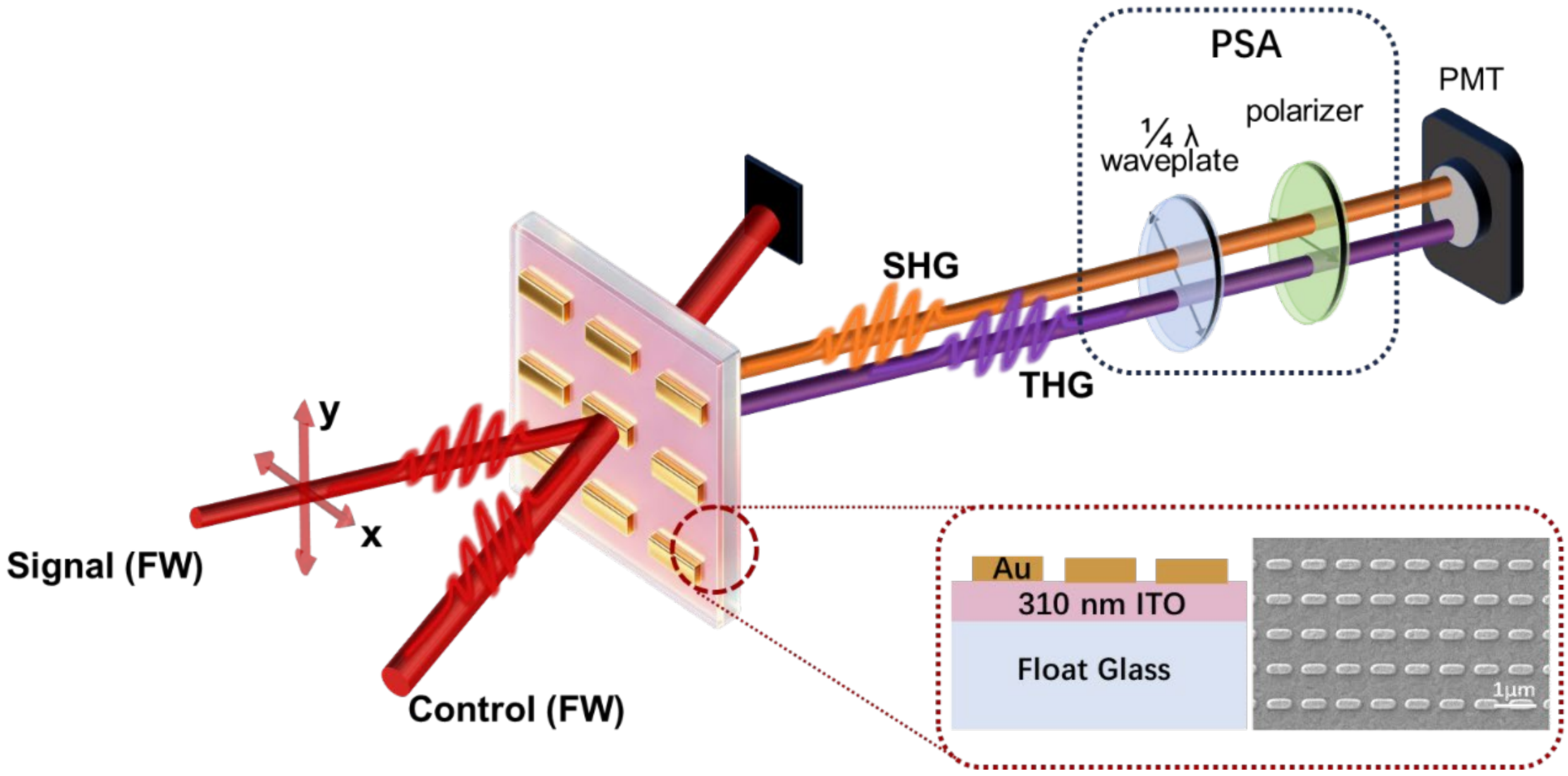

**FIG. 1** Schematic of femtosecond polarization modulation of broadband harmonic generation. In this setup, the SHG and THG generated by the FW signal light were recorded by a polarization state analyzer (PSA) and a photomultiplier tube system for polarization estimation; meanwhile, the transmitted FW signal light was filtered. The control light is always kept at the resonant fundamental wavelength and illuminated at $x$-polarization. Note: the $x$-polarized direction is parallel to the long axis of antennas. The red panel displays the cross-sectional diagram and the scanning electron microscopy image of the fabricated sample.

To transiently modify the polarization state of harmonic generation from signal light, an ultrafast anisotropic change in refractive index ($\Delta n_x(t) \neq \Delta n_y(t)$) should be induced in strongly coupled system by applying a control optical pulse. This is possible because the strongly coupled system in our study holds in-plane optical anisotropy through symmetry breaking of the coupled structure. Such a transient anisotropic nonlinear refractive index modifies the amplitude/phase of the orthogonal polarized components of harmonics differently, thereby enabling an ultrafast control of the complete polarization state of harmonic generation. Moreover, because SHG and THG originate from the same coupled system and share the same ultrafast anisotropic nonlinear response, their polarization states can be simultaneously modulated, establishing a general strategy for broadband, multi-wavelength ultrafast polarization control from the visible to UV region.

## Linear and nonlinear optical properties of anisotropic strong coupling

To realize strong coupling, the plasmonic resonant wavelength and its linewidth are optimized to match the ENZ mode of the ITO film by finely tuning the geometric parameters of the gold nanoantenna array (fabrication details of the sample are provided in the Methods section). In our case, the ENZ wavelength of ITO appears at 1240 nm; meanwhile, a 310-nm-thick ITO film was selected to increase the coupling strength for obtaining strong enhancement of local electric field[25]. For *x*-polarization, a coupled resonance centered at 1220 nm was clearly observed in the transmittance spectrum at normal incidence (**FIG. 2a**); another coupled resonance at a longer wavelength (~1900 nm) was suppressed by the strong metallic characteristics of the thick ITO film (**FIG. S1a**). Such a spectral behavior keeps consistent with the simulations and the previous study (**FIG. 2a** and **FIG. 2b**)[25]. Whereas, no strong coupling happens for the *y*-polarized illumination, indicating the anisotropic property of the coupled system (**FIG. S1b** and **FIG. S1c**). Although the coupled resonance shows a small spectral shift when changing the angle of the illuminated light, the anisotropic behavior of such strong coupling is still maintained.

In order to optimize the excitation conditions of harmonic generation for the polarization measurements below, without control light, only *x*- and *y*-polarized FW signal light (resonant excitation at 1220 nm) were applied to characterize the angle-dependent SHG and THG on the coupled system. As illustrated in **FIG. 2c**, remarkable SHG and THG peaks can be simultaneously observed at 407 nm and 610 nm, respectively, by choosing the specific excitation condition. Such harmonic generation spans UV to visible range, ensuring the broadband, multi-wavelength polarization control. The simulated field distributions at the fundamental and harmonic frequencies show that the enhanced fields are concentrated near the gold/ITO interfaces and extend significantly into the ITO layer (**FIG. S2a**). This spatial overlap suggests that the ITO layer plays an important role in the observed harmonic generation. For SHG, the harmonic process strictly relies on the out-of-plane symmetry breaking at the interface, necessitating a surface-normal electric field[26-28]. At oblique incidence, the SHG enhancement for *x*-polarized light is much larger than that for *y*-polarized light (**FIG. 2d**), indicating the role of strong coupling in enhancing second-order nonlinearity[29]. In the case of THG, its intensity under *x*-polarized excitation increases remarkably when increasing the incident angle. This enhancement arises from the emergence of an out-of-plane electric-field component at oblique incidence, which is amplified by the ENZ response of ITO, thereby strengthening the longitudinal fundamental field within the ITO layer and enhancing THG[30,31]. Surprisingly, the THG signal under *y*-polarized excitation is stronger than that under *x*-polarized excitation; this may come from the strong volume integral of the local fundamental electric field of *y*-polarization for the coupled system. As shown in Supplementary Note S2, the THG signal at the coupled resonance is dominated by ITO. Simulated results show that the integrated field contribution in ITO for *y*-polarization is larger than that for *x*-polarization over the entire range of incidence angles (**FIG. S2b**). This can be evidenced by THG measurements of pure 310-nm ITO film at orthogonal polarized excitations. The THG signal under *y*-polarized excitation for pure ITO film is larger than that under *x*-polarized excitation at oblique incidence (**FIG. S2c**).

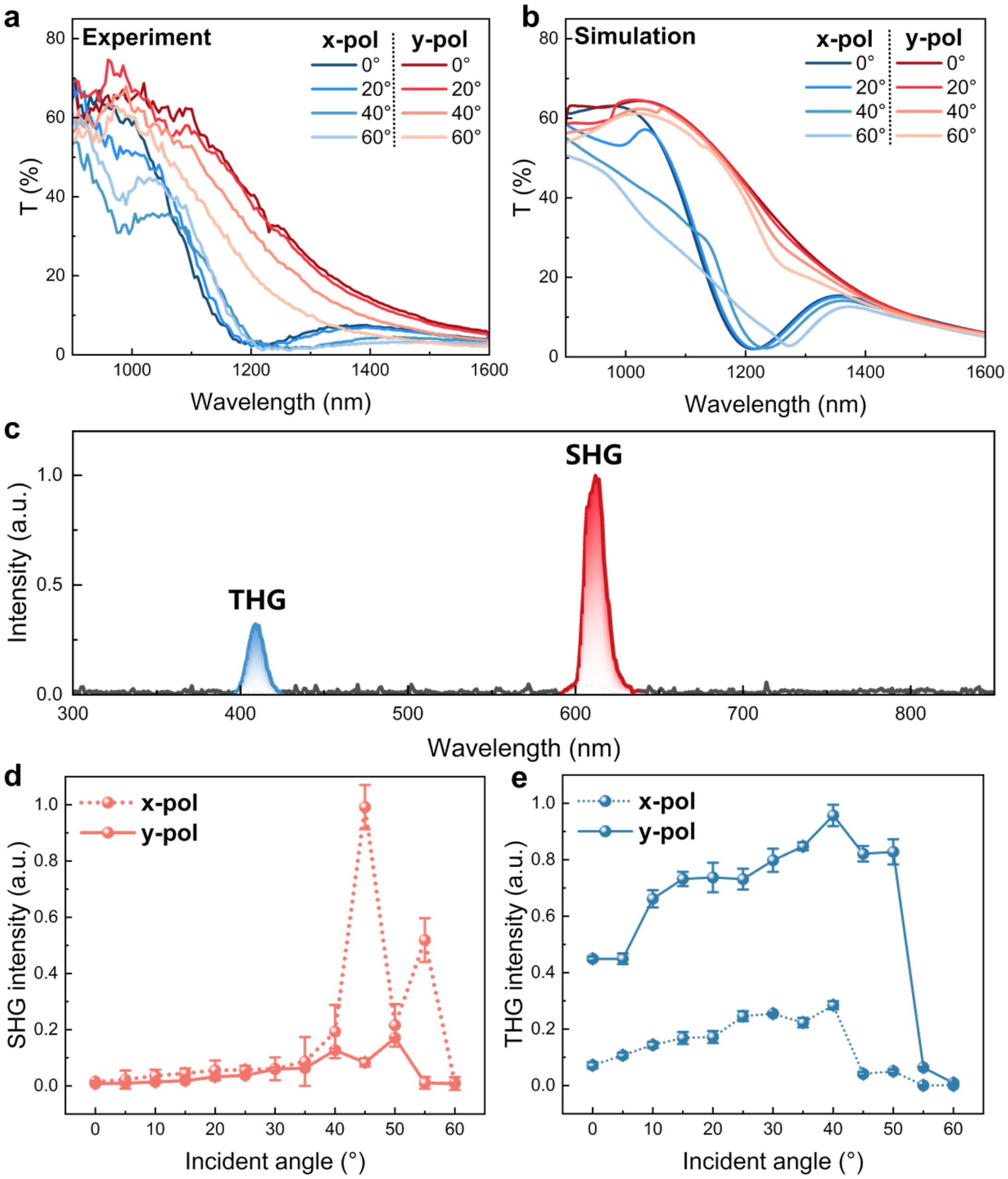

**FIG. 2 Linear and nonlinear anisotropic optical responses of the coupled system. a**, **b**, Experimental and simulated transmission spectra at different incident angles under *x*- and *y*- polarizations. **c**, Measured SHG and THG signals at nonlinear spectrum by a 1220-nm, p-polarized excitation with an irradiated intensity of 20 GW/cm². **d**, **e**, Incident angle-dependent SHG and THG intensities driven by the orthogonal polarized 1220-nm FW resonant excitations.

## Femtosecond anisotropic modulation of harmonics

To achieve transient anisotropic modulation of harmonic generation for ultrafast polarization control, we performed ultrafast amplitude modulation of SHG and THG from FW signal light at orthogonal polarizations under resonant pumping at 1220 nm. The amplitude modulation of the harmonics at time delay *t* (time interval between the control and signal pulses) is defined as the relative change of transmitted harmonic intensities with respect to their values without control light ( $I_{SHG_0}, I_{THG_0}$ ): $\frac{\Delta I_{SHG}(t)}{I_{SHG_0}} = \frac{I_{SHG}(t)-I_{SHG_0}}{I_{SHG_0}}, \frac{\Delta I_{THG}(t)}{I_{THG_0}} = \frac{I_{THG}(t)-I_{THG_0}}{I_{THG_0}}$ . The modulation depth corresponds to the maximum variation of the harmonic intensity. As presented in **FIG. 3a** and **FIG. 3b**, the transmittance of harmonics experiences rapid decreases, then recovers to their original values within ~600 fs. Specifically, the modulation depth of both SHG and THG for $x$-polarized FW signal light is clearly larger than that for $y$-polarized FW signal light when exciting the coupled system with the same control light. This behavior indicates the pump-induced femtosecond anisotropic modulation of harmonics in our coupled system. Such polarization-dependent modulation arises from the selective excitation of the coupled resonance. With the resonant excitation of control light, the nonlinear refractive index under strong coupling at $x$-polarized direction ($n_{2,x}$) is remarkably larger than that of uncoupled condition ($n_{2,y}$) by introducing strong local-field enhancement[13,14]. Therefore, the change in refractive index, which determines the modulation depth of harmonic fields, also holds the same anisotropy of $\Delta n_x > \Delta n_y$ when fixing the intensity of the incident control light[32,33]. Such an anisotropic modulation ensures the polarization control of harmonics.

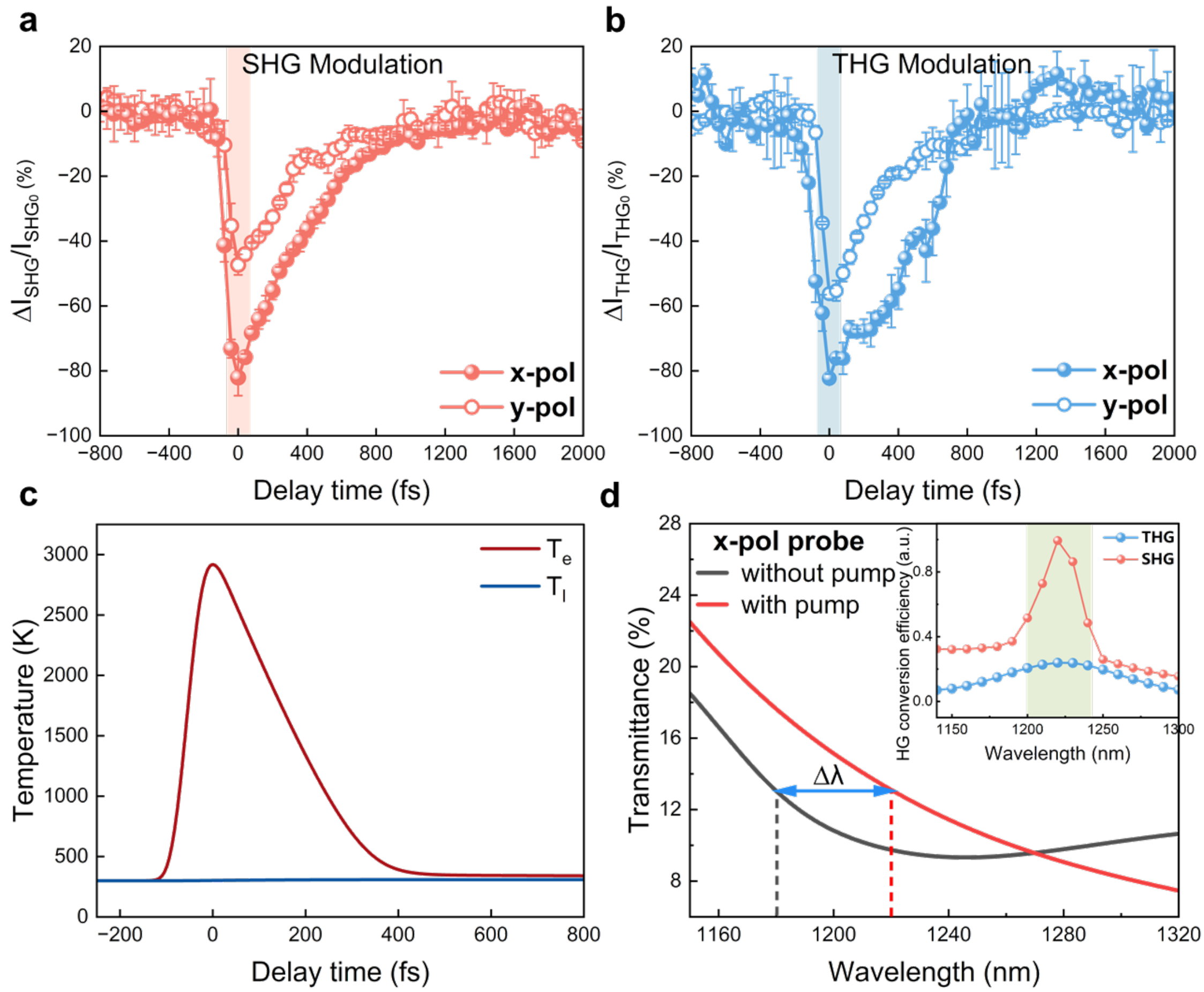

**FIG. 3 Femtosecond anisotropic nonlinear response of harmonic fields. a**, **b**, Femtosecond modulation of the transmitted SHG (a) and THG (b) intensities for *x*- and *y*-polarized FW signal light under a resonant excitation at 1220 nm with a fluence of 7 mJ/cm². **c**, Temporal evolution of the electron and lattice temperatures in ITO film with a control fluence of 7 mJ/cm². **d**, Transmittance spectra of the strongly coupled system for *x*-polarized signal light without pump (black curve) and with pump at zero-delay (red curve). Inset: Simulated wavelength-dependent conversion efficiencies of SHG (red curve) and THG (blue curve) under *x*-polarization with a FW intensity of 20 GW/cm². Here the green shaded area denotes the wavelength range of the coupled resonance.

To elucidate the temporal response of the coupled system, we consider the contribution from the ultrafast response of both the ITO film and gold antennas. These transient responses are determined by the energy redistribution processes of electrons in the conduction band upon excitation of the optical control pulse, and can be described by a two-temperature model[34-36] (See Supplementary Note S3). In this model, the optical energy of the control pulse is first stored in nonthermal electrons and then transferred to thermal electrons (Fermi distribution) via electron-electron scattering. After reaching the highest electron temperature, the hot electrons interact with phonons and transfer energy to the lattice system via an electron-phonon coupling process. Eventually, the electron and lattice reach a thermal equilibrium state. In the ITO film, the large electron-phonon coupling effect enables an ~400-fs effective relaxation process (**FIG. 3c**). In contrast, the relaxation of photoexcited hot electrons in gold antennas usually exhibits a relatively longer response time, i.e., 1-2 ps[37], due to the relatively weak electron-phonon coupling. Although the contribution from the gold nanoantenna slightly slows down the temporal response of the whole system, the coupled system still maintains a femtosecond modulation time (**FIG. 3a** and **FIG. 3b**)

To reveal the sharp decrease of SHG and THG signal under the control excitation, the transmission spectra of the coupled system were simulated with and without pumping. As shown in **FIG. 3d**, pump excitation (intra-band) induces a clear spectral redshift of the coupled resonance, such a spectral shift reaches ~40 nm under a control fluence of 7 mJ/cm$^2$. This can be understood by analyzing the pump-induced modifications of the permittivity of ITO at the ENZ band. With the excitation of control light, the elevated electron temperature leads to a reduction of the plasma frequency ($\omega_p$) of ITO, enabling a redshift of ENZ point[38,39]. The change of ENZ wavelength pushes the strong coupling to longer wavelengths, thereby detuning the coupled resonance from the excitation wavelength (See Supplementary Material S4). Consequently, the wavelength of control light moves away from the spectral region of the strongly coupled resonance (green shaded area in inset of **FIG. 3d**), leading to a pronounced decrease in the local field at the fundamental frequency, which gives rise to a pronounced suppression of harmonic conversion. This pump-induced detuning therefore accounts for the simultaneous decrease in both SHG and THG intensities, demonstrating that the ultrafast harmonic modulation arises from the photoinduced redshift of the strongly coupled resonance.

## Femtosecond polarization modulation of broadband harmonics

Considering the active modulation of broadband harmonics, the polarization state of SHG and THG light should be measured under the same excitation conditions. To estimate the polarization of harmonic generation, a polarization state analyzer (PSA) was introduced to record the Stokes parameters (**FIG. 1**). In order to maximize the nonlinear optical anisotropy and ensure the measurable harmonic light at different polarizations, the incidence angle of the FW signal light is selected to be 45°, and the intersection angle between the control and the FW signal light is fixed at 15°. For SHG modulation, we choose an azimuthal angle of 70° (with respect to $x$-polarized direction) for the incident FW signal light. Before the excitation of control light, the azimuthal angle of the transmitted SHG is -12.4°. After the arrival of the control pulse, the polarization ellipse of SHG undergoes a rotation of 55.0° and reaches -67.4° at zero delay, then recovers to its original value within ~600 fs (**FIG. 4a**). In the case of controlling THG, an incident azimuthal angle of 20° was used, the polarization ellipse shows a remarkable rotation from 36.9° to 13.6° within a similar time duration with respect to that of SHG modulation (**FIG. 4b**). These femtosecond polarization rotations originate from the polarization-dependent refractive index change of the anisotropic coupled system. Such a transient anisotropy changes both the amplitude ratio and phase difference between two orthogonal harmonic-field components, thereby rotating and deforming the polarization ellipse.

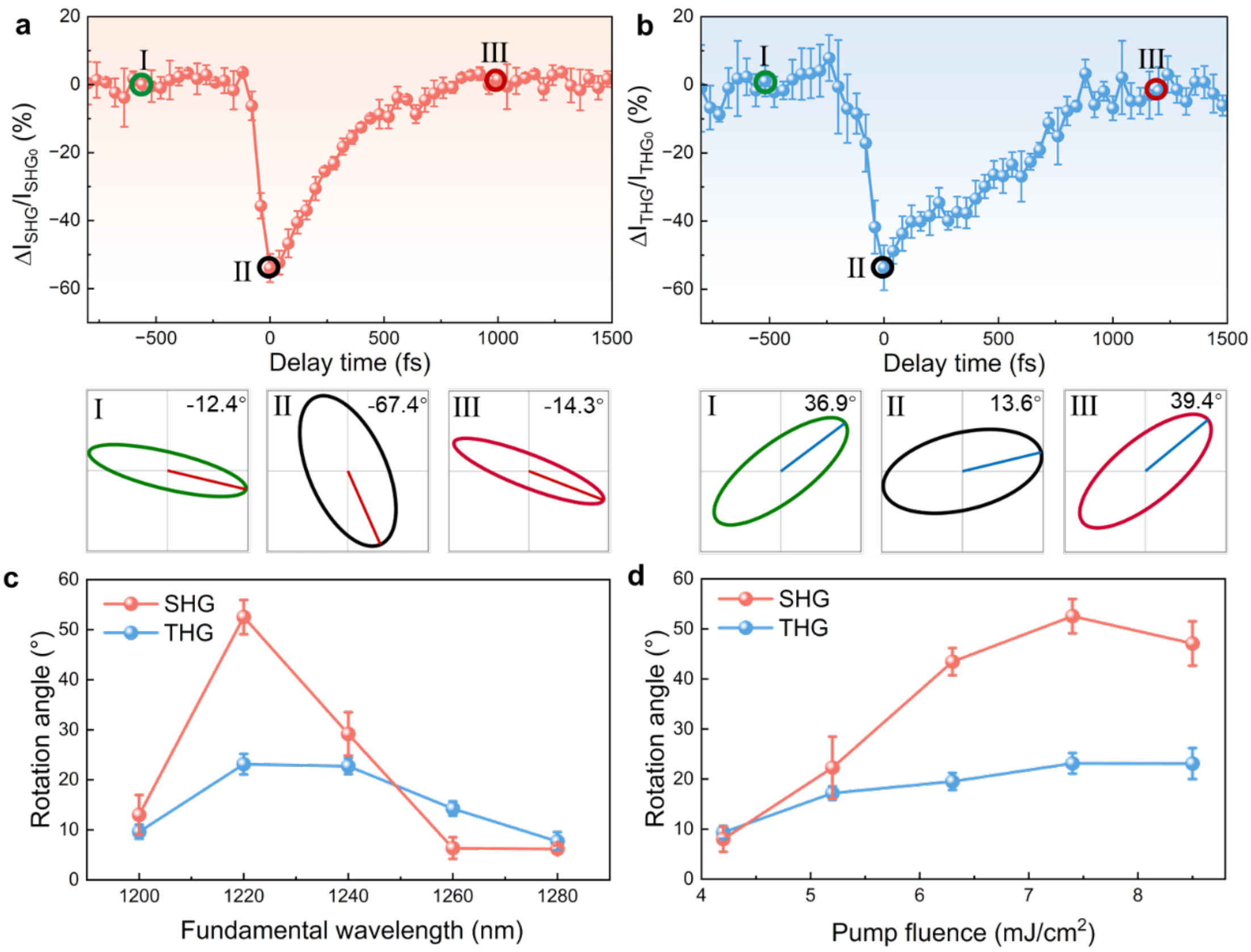

**FIG. 4. Femtosecond polarization modulation of harmonic generation. a**, Femtosecond intensity modulation and the corresponding polarization variation (bottom panels) of the SHG signal in the coupled system with an incident FW polarization azimuthal angle of 70° (defining *x*-polarization as 0°) under a pump fluence of 7 mJ/cm$^2$. **b**, Femtosecond evolution of the polarization state of THG signal for an incident FW polarization azimuthal angle of 20° under a control fluence of 7 mJ/cm$^2$. **c**, **d**, Maximum modulation of the polarization ellipse of the SHG and THG light as a function of wavelength of the incident FW signal light and pump fluence.

To further understand the properties of the polarization ellipse rotation of the harmonic generation, we also investigate the FW-wavelength-dependent and control-fluence-dependent polarization control at zero delay. As indicated in **FIG. 4c**, the rotation of the polarization azimuthal angle reaches a maximum value at strongly coupled resonance for both SHG and THG light. This verifies the critical role of anisotropic strong coupling on the harmonic polarization modulation. When changing the fluence of the control light, the modulation of the polarization ellipse shows strong fluence-dependent behavior, e.g., the rotation of the azimuthal angle for SHG increases as a function of fluence and reaches a maximum of 55.0° at 7.4 $mJ/cm^2$, followed by saturation. Whereas the rotation of the azimuthal angle for THG increases monotonically and achieves 25.3° at 8.5 $mJ/cm^2$ (**FIG. 4d**). These dependences demonstrate that the large polarization ellipse rotation of harmonics is not merely a consequence of pump-induced intensity modulation, but also closely associated with the resonance-enhanced anisotropic nonlinear response arising from the coupling between the plasmonic and ENZ mode.

## Cascaded femtosecond modulation of harmonic polarization

Knowing that the femtosecond efficient control of the polarization of signal light at the resonant fundamental wavelength has been investigated in ENZ-based systems[10,11], such a transient FW polarization control at the NIR wavelength can be transferred into ultrafast harmonic polarization modulation in the UV range by designing a cascaded all-optical polarization-modulation scheme, as sketched in **FIG. 5a**. In this configuration, the polarization of FW signal light at NIR wavelength can be transiently modulated in the strongly coupled metasurface (Sample A, the same as the sample used above) with the resonant excitation of control light; then the modulated transmitted FW signal light can generate a modulated THG light at the UV wavelength when passing through a 310-nm ITO film (Sample B). Therefore, Sample A acts as a transient polarization modulator of the FW signal light, while Sample B serves as an isotropic THG converter, which retains the information of polarization modulation. To maximize the change of the FW signal light polarization, Sample A should work at the largest anisotropic amplitude modulation condition (**FIG. S5**). Herein, the incident angle of the FW signal light is 30°, and the intersection angle between signal and control light is 15°. When the polarization azimuth angle of the incident FW signal light is 45°, the polarization ellipse of the transmitted FW signal light undergoes a 32.9° rotation within ~600 fs under a pump fluence of 7.2 mJ/cm$^2$ (**FIG. 5b**). Such a transient polarization control can be efficiently transferred into the UV region via THG by near-normal incidence on an ITO film. As shown in **FIG. 5c**, the polarization ellipse of the converted THG experiences a femtosecond rotation of 35.6°, which nearly maintains the polarization modulation information of the FW signal light. In addition, the full temporal evolution of the polarization state can be mapped onto a Poincaré sphere, where the trajectory captures both the azimuthal rotation and the accompanying change in ellipticity during the excitation and relaxation (**FIG. 5c**).

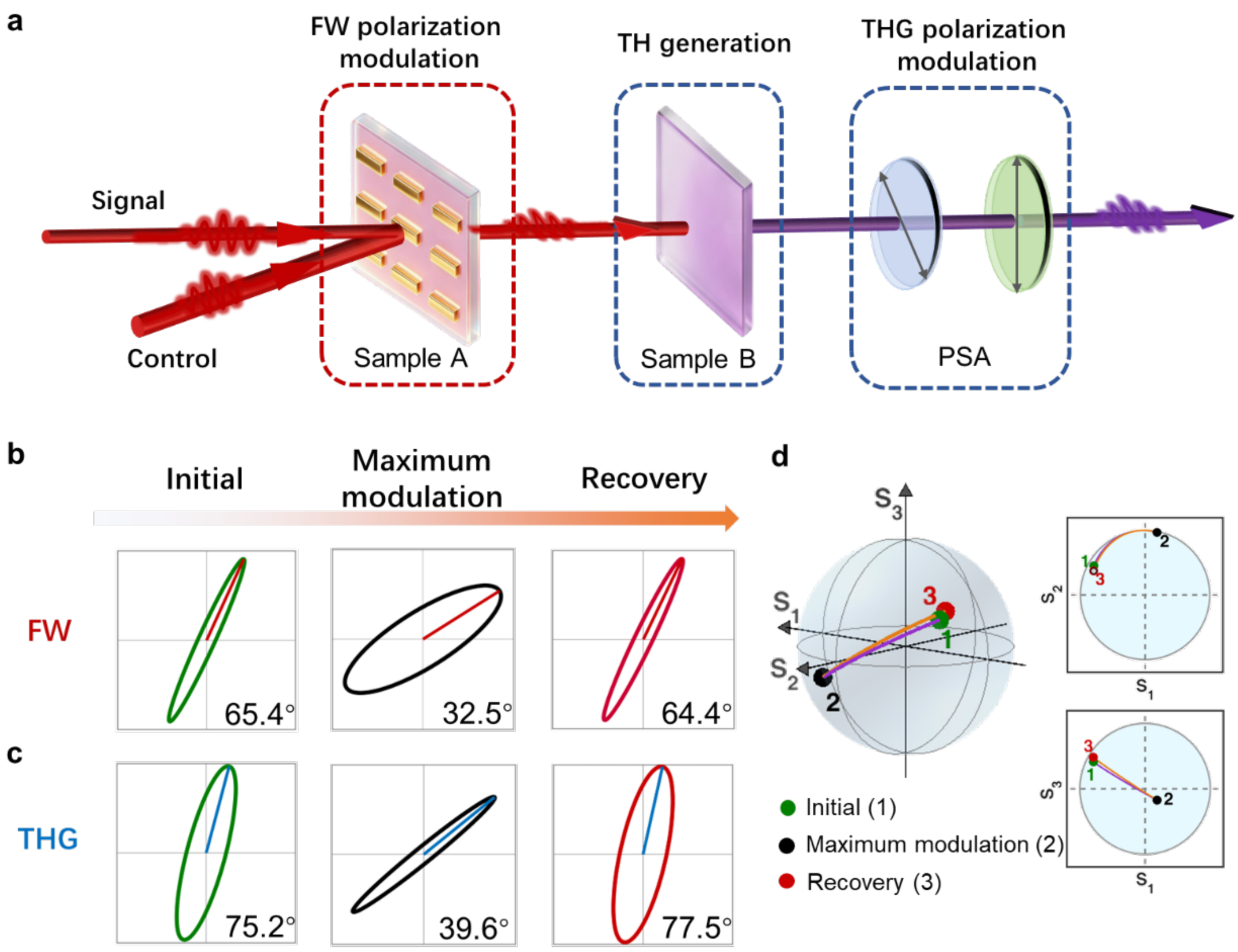

**FIG. 5 Cascaded all-optical femtosecond modulation of harmonic polarization. a**, Schematic of the cascaded femtosecond polarization modulation configuration. Sample A is the anisotropic strongly coupled metasurface, and Sample B is a 310-nm-thick ITO film. **b**, Pump-induced temporal evolution of the polarization ellipses of the FW signal light. **c**, Femtosecond modulation of the converted THG signal light. **d**, Evolution of the polarization state on the Poincaré sphere during modulation. The right panels show the corresponding projections onto the $S_1 \to S_3$ and $S_1 \to S_2$ planes.

# Discussion

The active polarization modulation strategies in this study achieve several key advances that collectively define a new paradigm for femtosecond nonlinear polarization control at short wavelengths. First, the operation wavelength range has been extended into UV region by means of THG. Meanwhile, a polarization elliptic rotation of >30° has been achieved at 407 nm. This modulation can be pushed to even shorter wavelengths (such as extreme UV) by designing a coupled metasurface with high-efficiency high-order harmonics[40-42]. Second, in contrast to the conventional ultrafast polarization control for the narrow-band, single resonant fundamental wavelength, the use of harmonics in our study facilitates the broadband and multiwavelength optical modulation in a single platform. The number of operating wavelengths of the transient polarization modulation can be even increased by considering other types of nonlinear wave-mixing effects, including fourth-harmonic generation, four-wave mixing, and sum-frequency generation[43]. By hybridizing structural anisotropy with time-induced dynamic anisotropy, our platform not only provides a universal architecture for tailoring transient polarization state, but also sheds light on broadband, ultrafast and high contrast-ratio polarization-related modulators with a nanoscale footprint. For the cascaded modulation strategy, the key point is to design a coupled system that exhibits efficient modulation for the resonant fundamental frequency. The modulation of FW light can be converted into the modulation of corresponding higher harmonic frequency, providing a route towards remotely programmable and spectrally translated ultrafast polarization modulation.

In conclusion, we have demonstrated a femtosecond, broadband light polarization modulation across UV to visible wavelength range in an anisotropic strongly coupled system. We achieved polarization ellipse rotation of 23.3° and 55.0° within 600 fs for SHG- and THG-signal light, respectively. The transient modulation of the harmonic polarization stems from hot electrons of the coupled structure with different thermalized temperatures at various polarizations. In addition, a cascaded femtosecond polarization modulation of UV light was also investigated by converting the transient polarization modification of the NIR fundamental light to that of THG. These strategies not only chart a new course for ultrafast nonlinear optics, but also open possibilities for high-speed optical information processing and programmable photonic chips.

# Methods

## Fabrication of the metasurface

The sample of anisotropic strong coupling consists of a gold nanoantenna array that is fabricated on top of a 310-nm-thick ITO film. The nanoantenna array was patterned by electron-beam lithography using a bilayer PMMA resist and subsequently formed by e-beam evaporation of a 90-nm-thick gold film, followed by a lift-off process. The dimensions of the gold antennas are 580 nm × 180 nm × 90 nm, with a periodicity of 800 nm. This ensures the spectral overlap of the localized plasmon resonance and the ENZ mode of ITO for optical strong coupling in the *x*-polarized direction.

## Numerical simulations

*Linear optical transmission simulations*

To investigate the anisotropic optical responses of the coupled system, full-wave numerical simulations were performed. In the linear regime, the coupled system was illuminated by plane waves that propagate along the positive *z*-axis. The incident angle and specific polarization states are configured via boundary conditions. Periodic boundary conditions were applied along the *x*- and *y*-axes to emulate an infinite two-dimensional array. Concurrently, perfectly matched layers were implemented at both ends of the *z*-axis to completely absorb the outgoing waves, thereby eliminating any non-physical artificial boundary reflections. The dielectric functions of gold[44] and $SiO_2$[45] were adopted from the reference paper. The dispersion of the ITO film was precisely modeled using a relative permittivity model to capture its ENZ characteristics around 1240 nm accurately.

*Nonlinear responses*

We employ a two-step coupled model for the nonlinear simulations of SHG and THG. Specifically, a secondary, independent electromagnetic-wave, frequency-domain interface was introduced, operating at the harmonic frequencies. The highly localized fundamental electric field, $E(\omega)$, extracted from the first step within the gold nanoantennas and the ITO film, was utilized to construct the macroscopic nonlinear polarization sources. The polarization vectors acted as the sole internal excitation sources driving the harmonic radiation. Broadband harmonic response spectra were obtained through wavelength-domain parametric sweeps, and the far-field harmonic conversion efficiencies were quantitatively determined by integrating the time-averaged Poynting vector over the transmission boundary.

## Optical measurement setup

*Linear optical characterization*

Linear transmission spectra of the coupled system were characterized by using an ultra-broadband tungsten halogen lamp (50W, LSTS-50). This lamp can provide a continuous-wave (CW) white-light source for the measurements. The output beam was spatially shaped, collimated, and focused onto the sample at normal incidence with a spot diameter of ~200 μm. The polarization of the incident light was precisely controlled to be either $x$- or $y$- polarized using a broadband linear polarizer. The transmitted light was subsequently recorded and spectrally resolved by a commercial monochromator (Omni-λ series, Zolix Instruments Co., Ltd.).

*Nonlinear harmonic measurements*

All the nonlinear optical measurements were conducted with a femtosecond laser system, which consists of a Ti: sapphire amplifier (Libra, Coherent) and an optical parametric amplifier (OPA). This system can provide wavelength-tunable femtosecond optical pulses in the NIR regime. To ensure that the measured harmonic fields from the FW signal light are generated by the sample, the residual visible stray light from the laser system was eliminated by leaving a long-pass filter (FGL1000, Thorlabs) at the output of the OPA. The FW signal light was focused onto the sample with a beam diameter of 80 μm. A combination of a half-wave plate (HWP) and a linear polarizer was utilized to continuously tune the polarization state and power of the incident FW signal light. After the sample, the bandpass filters were applied to block the transmitted FW signal light. The transmitted harmonic fields were eventually recorded by a monochromator and a photomultiplier tube (PMT, H11902-20, Hamamatsu) system.

*Time-resolved harmonic polarimetry*

To investigate the femtosecond active control of harmonic polarization, a non-collinear pump-probe configuration was constructed. The output optical beam from the OPA was split into two portions. One was used as control light (pump), and the other served as the FW signal light. A motorized delay line was placed into the control path to tune the time delay between the two pulses. After exciting the sample, the transmitted control light was blocked. In our case, the intersection angle between the control beam and the FW signal beam is 15°. For the incident FW signal light, its intensity and polarization can be tuned by a HWP and a linear polarizer. The signal light was focused onto the sample with a spot diameter of 80 μm (measured at normal incidence). After passing through the sample, the transmitted FW signal light was filtered, whereas only harmonic fields from the sample were sent to the customized PSA system for estimation of Stokes parameters. In our study, the PSA consists of a rotatable quarter-wave plate (QWP) followed by a fixed linear polarizer (with its transmission axis aligned horizontally). For a given fast-axis angle $\beta$ of the QWP, the transmitted harmonic intensity $I_T(\beta,\delta)$as a function of the Stokes parameters $(S_0, S_1, S_2, S_3)$ is given by[46]:

$$I_T(\beta,\delta) = \frac{1}{2}\{S_0 + (S_1 cos2\beta + S_2 sin2\beta)cos\beta \\ - [(S_2 cos2\beta - S_1 sin2\beta)cos\delta + S_3 sin\delta]sin2\beta\}$$

To improve the polarimetric accuracy, the transmitted intensities were recorded at eight discrete QWP angles (0°、15°、30°、45°、60°、75°、90° and 105°). This overdetermined system of eight equations was then solved to uniquely extract the four unknown Stokes parameters at each time delay.

*Cascaded all-optical modulation of harmonic polarization*

The cascaded optical modulation of harmonic polarization setup consists of two samples. The cascaded modulation setup resembles the pump-probe configuration of the direct harmonic modulation above. The control light and the FW signal light were spatially overlapped on sample A with a relatively larger spot diameter of 300 μm. The main difference is that the harmonic generation from the FW signal light after sample A was filtered, whereas the transmitted FW signal light after sample A was kept as the incident light source for sample B (ITO film). The polarization modulation of the FW signal light was detected by a NIR PSA and a biased InGaAs photodetector (DET10N2, Thorlabs). After converting the transmitted FW signal light into harmonic signal light by sample B, the polarization state of the harmonic signal light was recorded by a UV PSA and PMT system.

## Acknowledgements

We acknowledge support from the National Natural Science Foundation of China under Grant No. 12304371 and the Young Talent Support Plan of Xi'an Jiaotong University.

## Author Contribution

K.W. conceived the idea and fabricated the sample; Y. Z. performed the simulations and conducted all the optical measurements and the corresponding data analysis with the help and guidance of K. W.; Y. Z. and K.W. did the theoretical part. All authors contributed to discussing and commenting on the manuscript; K.W. supervised the project.

## Conflicts of Interest

There are no conflicts to declare.

## Data Availability Statement

The data that support the findings of this study are available from the corresponding author upon reasonable request.

## Supplemental document

See Supplement for supporting content.

Supplementary Information for

# Femtosecond all-optical polarization manipulation of harmonic generation in an anisotropic strongly coupled plasmon-epsilon-near-zero metasurface

Yingying Zhang, Jinhai Si, Xun Hou, Kuidong Wang*

*Key Laboratory for Physical Electronics and Devices of the Ministry of Education and Shaanxi Key Lab of Information Photonic Technique, School of Electronic Science and Engineering, Xi'an Jiaotong University, Xi'an, 710049, China*

Email: wangkuidong@xjtu.edu.cn

## Supplementary Note S1. Anisotropic electric field enhancement in the coupled system

To elucidate the strong-coupling characteristics of the sample, we simulated the transmission spectra of the coupled structure and the bare 310-nm-thick ITO film. The upper coupled resonance is located at 1220 nm, whereas the lower coupled resonance is not spectrally resolved because of the high optical loss of the ENZ film[1] (**Figure S1a**).

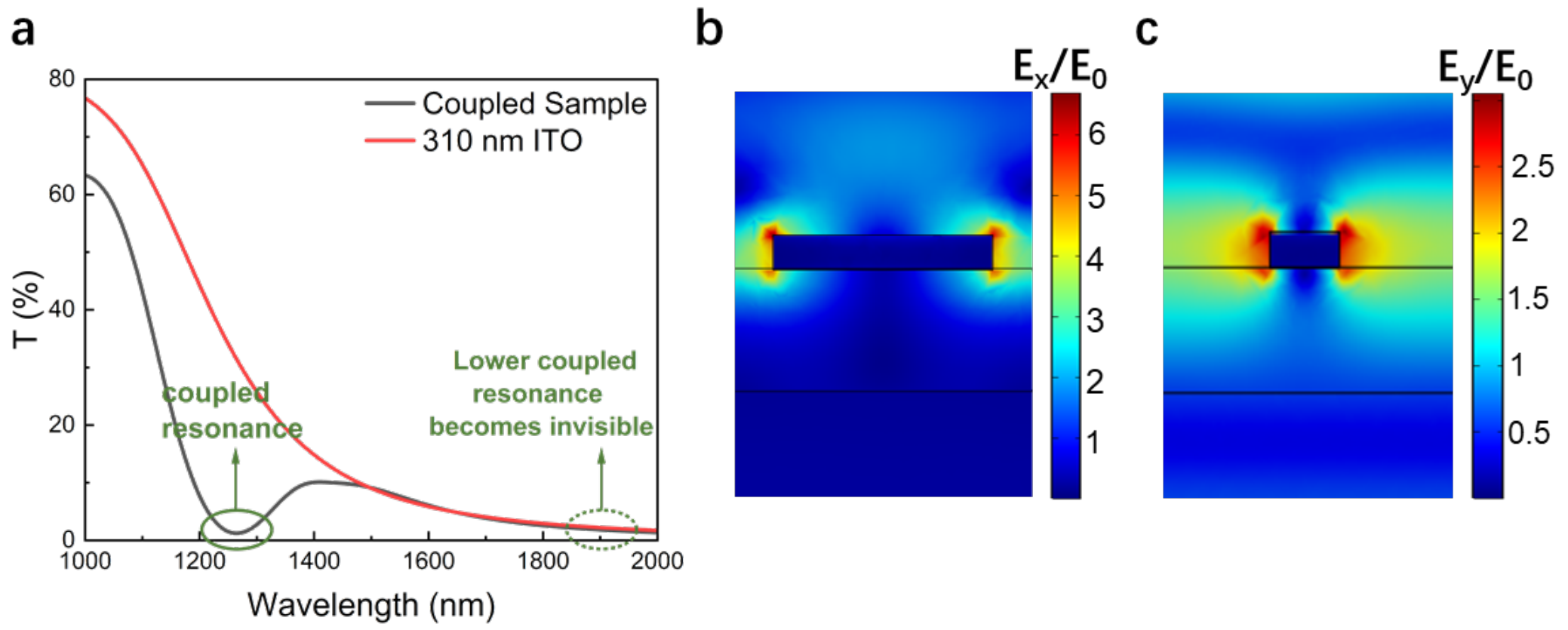

**Supplementary Figure S1. a.** The simulated transmittance spectra of the coupled system (black curve) and the reference sample (310-nm-thick ITO film, red curve) with *x*-polarized light at normal incidence. **b. c.** The electric field profiles under normally incident *x*- and *y*-polarized excitation, respectively.

The electric-field distributions of the coupled system at resonance indicate that the field enhancement for *x*-polarized excitation is clearly larger than that for *y*-polarized excitation (**Figure S1b** and **S1c**). This highlights the large optical anisotropy of the coupled system. the harmonic near-field hotspots are predominantly localized at both ends of the gold nanoantennas and near the gold-ITO interfaces, with most of the enhanced field confined within the ITO layer. This spatial localization indicates that the ITO layer plays a major role in the harmonic generation.

## Supplementary Note S2. Angular selectivity of THG from ITO film

The simulations were performed for the ITO structure at the fundamental resonant wavelength of 1220 nm. The spatially integrated third-order nonlinear polarization in the ITO layer was used as an indicator for the THG intensity. For an isotropic third-order response, the nonlinear polarization is given by $P^{(3)}(3\omega) = \varepsilon_0\chi^{(3)}E^3(\omega)$; consequently, the volume-integrated electric-field THG intensity can be approximated as $I_{THG} \propto \int_{ITO}|E(\omega)|^6\, dV$. The calculated integral for *y*-polarization is consistently larger than that for *x*- polarization at the oblique incident angle (20°~50°). This result indicates that, although *y*-polarized light does not excite the conventional normal ENZ field, its in-plane electric field distributed over a larger effective volume of the ITO layer can generate a strong third-order nonlinear polarization. The calculated $x$-polarized response increases with incidence angle and reaches a maximum near 35°, which consistent with enhanced ENZ coupling through the normal electric-field component at oblique incidence. Such a polarization-dependent field enhancement can be further understood from the angle-resolved THG signal of a pure ITO film with two orthogonal incident polarizations. Over most of the investigated angular range, the THG intensity under $y$-polarized excitation is larger than that under $x$-polarized excitation, whereas the THG generated by $x$-polarized excitation generally increased with increasing incidence angle. This agrees with the simulation, suggesting that the ITO layer plays key roles in the generation of third-harmonic light in our coupled sample.

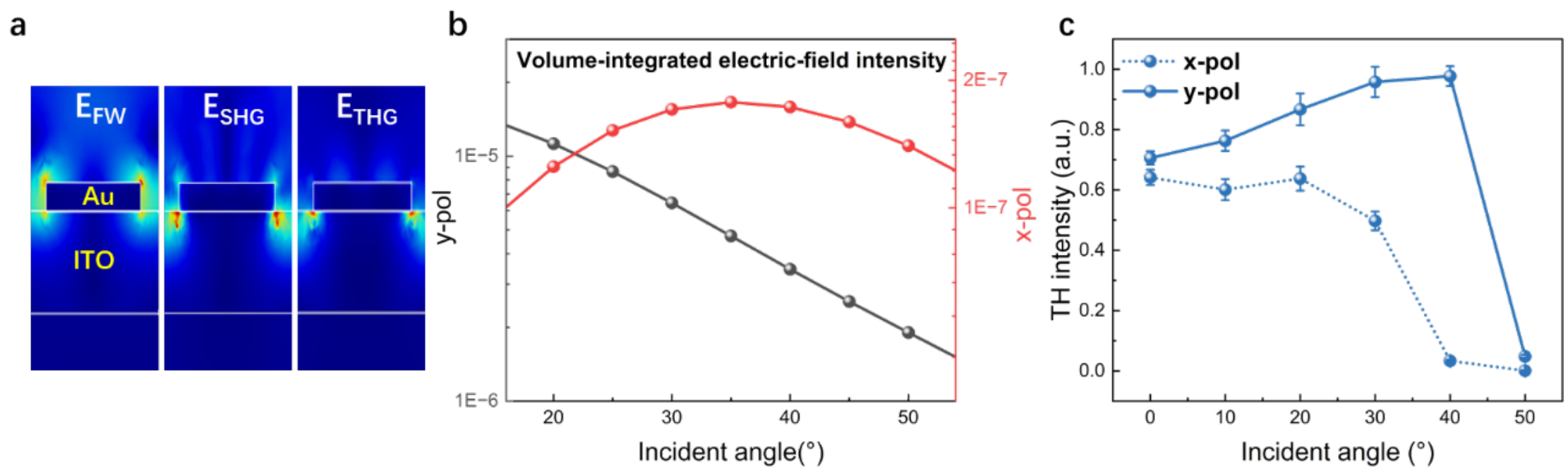

**Supplementary Figure S2. a.** Electric-field distributions of the resonant fundamental field, second-harmonic and third-harmonic fields under *x*-pol. at normal incidence. **b.** Angle-dependent volume-integrated electric-field response of the ITO film under *x*- and *y*-polarized excitation. **c.** THG intensity measured at different incidence angle for 310- nm-thick ITO film under *x*- and *y*-polarized fundamental light excitation.

## Supplementary Note S3. Temporal response in strongly coupled system

The temporal response of the coupled system at resonant pumping arises from the ultrafast dynamics of both ITO layer and gold nanoantennas. The ultrafast response of ITO originates from the change of the energy distribution of electrons in the conduction band under laser-induced electron thermalization. The temporal response of the hot electrons under intra-band pumping can be described by a two-temperature model. In this model, the optical energy from the control pulse is first stored in nonthermal electrons, then transferred to thermal electrons via electron-electron scattering. After that, the hot electrons interact with the lattice system via electron-phonon coupling process, and eventually they reach a thermal equilibrium. The dynamics can be related in terms of the nonthermal electronic energy density $N_{ITO}$, thermal electron temperature $T_{e,ITO}$, and lattice temperature $T_{l,ITO}$. These time-dependent processes can be described by[2,3]:

$$\frac{dN_{ITO}}{dt} = Q_{ITO} - (\gamma_{ee} + \gamma_{ep})N_{ITO} \tag{1}$$

$$C_e \frac{dT_{e,ITO}}{dt} = -G_{ITO}(T_{e,ITO} - T_{l,ITO}) + \gamma_{ee}N_{ITO} \tag{2}$$

$$C_l \frac{dT_{l,ITO}}{dt} = G_{ITO}\left(T_{e,ITO} - T_{l,ITO}\right) + \gamma_{ep} N_{ITO} \quad (3)$$

Here, $Q_{ITO}$ denotes the absorbed pump energy density, the coefficient of electron-electron scattering $\gamma_{ee} = 1.08 \times 10^{14}$, the energy transfer from the hot electrons towards the lattice $\gamma_{ep} = 3.31 \times 10^{13}$, $G_{ITO}$ is the electron-phonon coupling factor, $C_e$ and $C_l$ are the electronic and lattice heat capacities of ITO, respectively. $C_l = 2.58 \times 10^6$ $Jm^{-3}K^{-1}$. $C_e = \beta T_e$, $\beta = 3.81$ $Jm^{-3}K^{-2}$. By considering the expressions of the coefficients above[4], and applying the control fluence of 7.0 $mJ/cm^2$ and pulse-width of 50 fs, the time-dependent electron and lattice temperatures can be calculated as shown in **Fig 3c.** in the main text. Both ITO and gold antenna contribute to the transient optical response of the coupled system. However, the plasmonic nonlinear response of the gold nanoantennas occurs on a much shorter timescale, whereas the hot-electron relaxation dynamics in ITO predominantly govern the recovery process. Therefore, the two-temperature model (TTM) of ITO can be employed to approximate the principal ultrafast temporal characteristics of the coupled system.

## Supplementary Note S4. Spectral shift induced by temperature-dependent plasma frequency of ITO

Pump excitation induces a pronounced spectral redshift of the resonance, as shown in **Fig 3d**, indicating a substantial transient modification of the permittivity of the coupled system. To elucidate the microscopic origin of this spectral evolution, we developed a model that connects the absorbed pump energy to hot-carrier dynamics, ultimately, to the transient optical response.

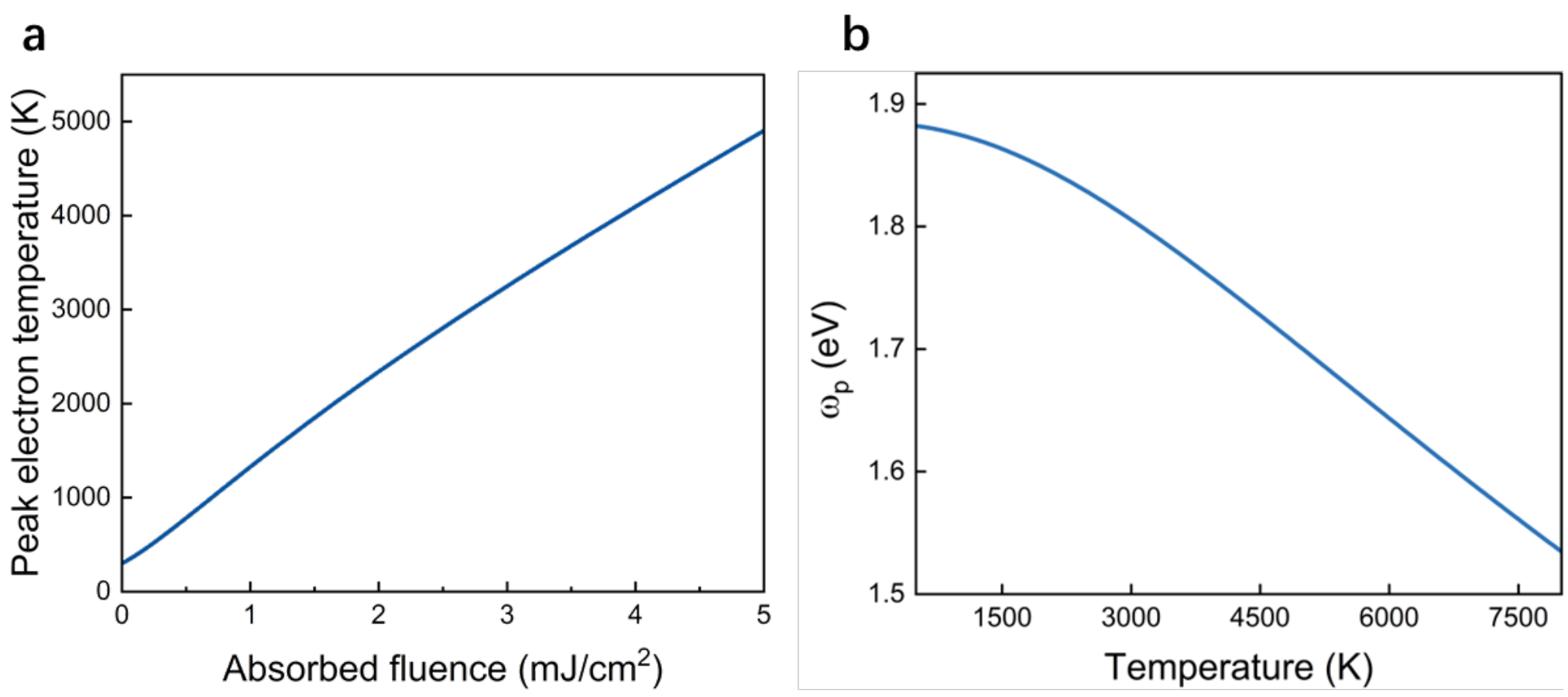

**Supplementary Figure S3. a.** Zero-delay transmittance modulation of the ITO film for 1220-nm pump-probe measurements at various absorbed pump fluences. **b.** Electron-temperature dependence of the plasma frequency.

The static electromagnetic response of the coupled system was first calculated using three-dimensional frequency-domain simulations. $A_{ITO}$ is the absorptance of the ITO layer for the pump light. According to the hot-electron dynamics of ITO, femtosecond excitation initially drives the absorbed energy into a nonthermal electron population, followed by electron thermalization and subsequent energy transfer to the lattice through electron-phonon coupling. By evaluating Eqs. (1)-(3), we obtained the results presented in **Fig. S3a**, which show a systematic increase in the electron temperature with increasing pump fluence. Electronic heating broadens the carrier distribution and redistributes electrons towards higher-energy states with larger effective masses, thereby reducing the Drude weight and the effective plasma frequency. The ultrafast optical response of ITO is strongly influenced by the non-parabolicity of its conduction band:

$$E(1+\alpha E)=\frac{\hbar^2 k^2}{2m_0^*} \tag{4}$$

Where $m_0^*$ =0.263$m_e$, $\alpha = 0.65$ eV$^{-1}$. The corresponding wavevector is

$$k(E)=\frac{\sqrt{2m_0^* E(1+\alpha E)}}{\hbar} \tag{5}$$

The free-carrier density was fixed at $n_e = 7.31 * 10^{26} m^{-3}$ . At each electron temperature, the chemical potential was determined by:

$$n_e=\int_0^\infty D(E) f(E,T_e)\, dE \tag{6}$$

$$f(E,T_e)=\frac{1}{1+exp\left[\frac{E-\mu(T_e)}{k_B T_e}\right]} \tag{7}$$

where $D(E)$ is the electronic density of states, $\mu$ is the chemical potential, and $E$ is measured from the conduction-band minimum. The temperature-dependent plasma frequency was subsequently evaluated as[5]:

$$\omega_p^2(T_e) = \frac{e^2}{3\pi^2\varepsilon_0 m_0^*}\int_0^\infty \frac{k^3(E)}{1+2\alpha E}(-\frac{\partial f}{\partial E})dE \quad (8)$$

$\varepsilon_0 = 8.85e^{-12}$ F/m, the thermally broadened carrier distribution induces a shift in the chemical potential to conserve the total electron density, while the increased occupation of higher-energy states with larger effective masses reduces the effective plasma frequency. This pump-induced modification of the plasma frequency, in turn, leads to a transient change in the dielectric permittivity of ITO. The complete simulation relation is therefore

$$A_{ITO} \to T_{e,ITO}(t) \to \omega_{p,ITO}(t) \to \varepsilon(\omega,t) \to T(\lambda,t) \quad (9)$$

Thus, pump excitation increases $T_{e,ITO}$, resulting in a reduction of $\omega_{p,ITO}$, while thermalization of the gold nanoantenna increases its Drude damping. These simultaneous perturbations alter the spectral position, linewidth and oscillator strength of the hybrid resonance. Consequently, our calculations reveal a spectral redshift of approximately 40 nm with a fluence of 7 mJ/cm$^2$.

## Supplementary Note S5. Femtosecond anisotropic modulation of fundamental-wave signal light

We investigated the ultrafast modulation response of the coupled sample in the coupled resonance regime (1220 nm) at an incidence angle of 30°. As shown in the figure, the modulation amplitude under *x*-polarized excitation is approximately 30 times larger than that under *y*-polarized excitation, revealing a pronounced polarization anisotropy. Despite the substantial difference in modulation strength, the two polarizations exhibit similar temporal dynamic.

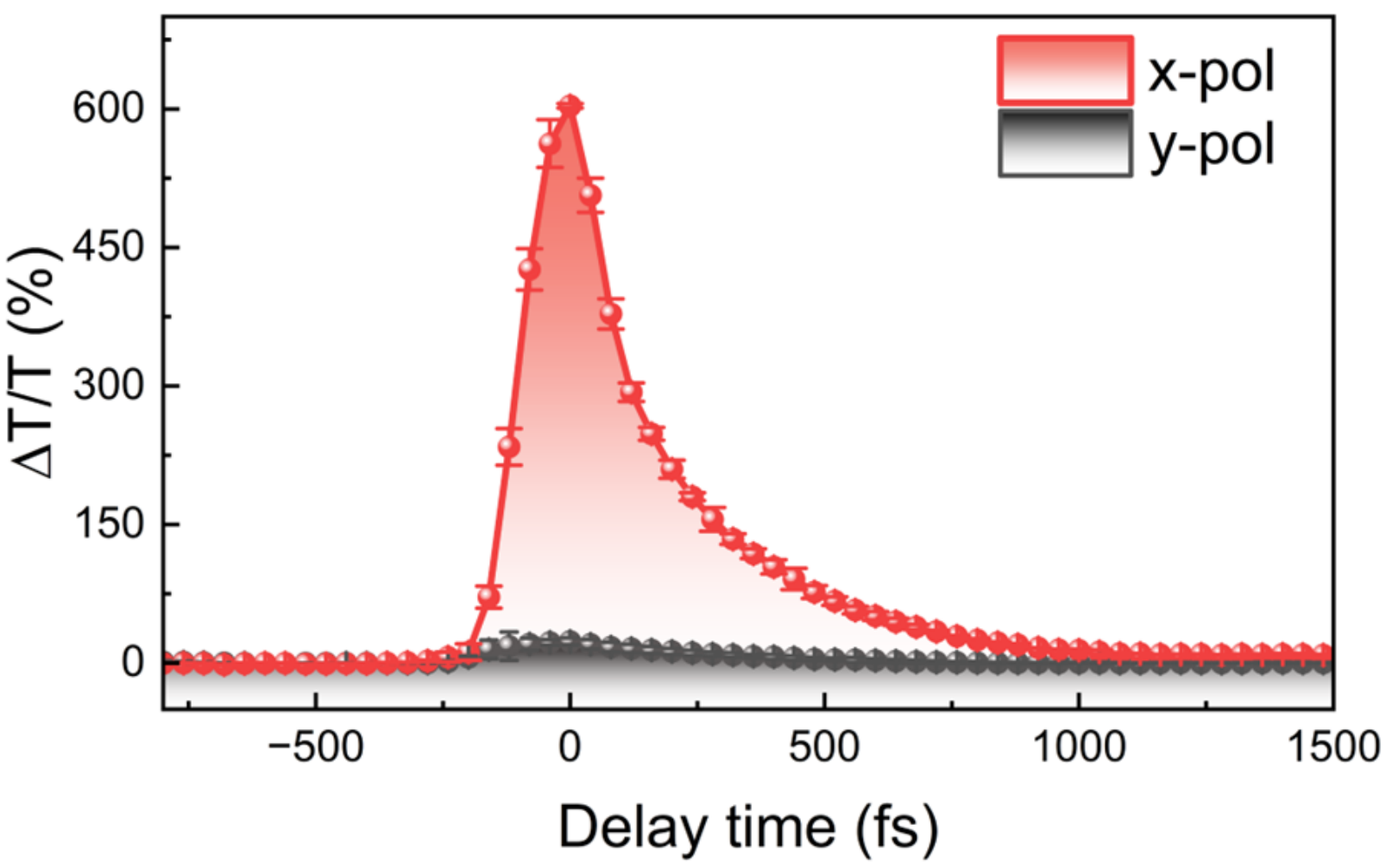

**Supplementary Figure S4.** Transient optical modulation of the coupled system for *x*- (red curve) and *y*-polarized (black curve) 1220 nm signal light under 1220 nm resonant excitation at 6 mJ/cm$^2$ irradiated fluence.